\documentclass[12pt]{article}
\usepackage{amsfonts}
\usepackage{graphicx}
\usepackage{epsfig}
\usepackage{amsmath}
\usepackage{amssymb}
\usepackage{multirow}
\begin{document}
\begin{titlepage}
\begin{flushright}
ZU-TH 22/12\\
CP3-12-38
\end{flushright}

\begin{center}
{\Large {\bf 
{Diphoton production in the ADD model to NLO+parton shower accuracy at the 
LHC }}} 
\\[1cm]

R. Frederix$^{a}$,
\hspace{.5cm} Manoj K.\ Mandal$^{b}$,
\hspace{.5cm} 
Prakash Mathews$^{c}$,
\hspace{.5cm} 
V. Ravindran$^{b}$,
\hspace{.5cm} 
Satyajit Seth$^{c}$,
\hspace{.5cm} 
P. Torrielli$^{a}$,
\hspace{.5cm} 
M. Zaro$^{d}$
\\[1cm]

${}^a$ Institut f\"ur Theoretische Physik, Universit\"at Z\"urich, 
Winterthurerstrasse 190, CH-8057 Z\"urich, Switzerland
\\[.5cm]

${}^b$ Regional Centre for Accelerator-based Particle Physics\\ 
Harish-Chandra Research Institute, Chhatnag Road, Jhunsi,\\
Allahabad 211 019, India
\\[.5cm]

${}^c$ Saha Institute of Nuclear Physics, 1/AF Bidhan Nagar, Kolkata 700 064, 
India
\\[.5cm]

${}^d$
Centre for Cosmology, Particle Physics and Phenomenology (CP3)\\
Universit\'e Catholique de Louvain\\
Chemin du Cyclotron 2, B–1348 Louvain-la-Neuve, Belgium
\\[.5cm]

\end{center}
\vspace{1cm}

\begin{abstract}
\noindent 
In this paper, we present the next-to-leading order predictions for
diphoton production in the ADD model, matched to the HERWIG parton
shower using the MC@NLO formalism. A selection of the results is 
presented for $d=2-6$ extra dimensions, using generic cuts as well 
as analysis cuts mimicking the search strategies as pursued by the 
ATLAS and CMS experiments.
\end{abstract}

\vspace{.7cm}

\end{titlepage}

\section{Introduction}

Extra dimension models \cite{ADD,RS1} are important new physics scenarios 
at the TeV scale that address the hierarchy problem and are being 
extensively studied at the LHC.  Recently both ATLAS \cite{atlas} and CMS 
\cite{cms} have looked for evidence of extra spatial dimension in the 
diphoton final state and put bounds on the fundamental Planck scale $M_S$
in $(4+d)$-dimensions, for the 7 TeV proton-proton collision.  

In the extra dimension models, Kaluza-Klein (KK) modes of a graviton
(as a result of the graviton propagating in the extra dimensions), could
decay to a pair of photons.  Interaction of the massive spin-2, KK modes
$h_{\mu \nu}^{(\vec n)}$ with the standard model (SM) particles localised
on a 3-brane, is {\em via} the energy momentum tensor $T^{\mu \nu}$ of 
the SM
\begin{eqnarray}
{\cal L} = - \frac{\kappa}{2} \sum_{(\vec n)} T^{\mu \nu} (x) 
h_{\mu \nu}^{(\vec n)} (x) ,
\end{eqnarray}
where $\kappa=\sqrt{16 \pi}/M_P$ is the reduced Planck mass in
4-dimensions.  In process involving virtual exchange of KK modes 
between the SM particles, like in the diphoton production process, 
the sum of the KK mode propagator ${\cal D} (s)$ is given by
\begin{eqnarray}
\kappa^2 {\cal D} (s) &=& \kappa^2 \sum_{n} \frac{1}{s-m_n^2+i 
\epsilon} ,
\nonumber \\
&=& \frac{8 \pi}{M_S^4} \left( \frac{\sqrt{s}}{M_S}\right )^{(d-2)}
\left[ -i \pi + 2 I \left(\frac{\Lambda}{\sqrt{s}}\right) \right] ,
\label{KKsum}
\end{eqnarray}
where $s$ is the partonic center of mass energy, $d$ is the compactified 
extra spatial dimensions, $\Lambda$ is the UV cutoff of the KK mode sum 
which is usually identified as $M_S$ \cite{HLZ,GRW} and the integral 
$I(\Lambda/Q)$ is given in 
\cite{HLZ}.  Note that in Eq.~\ref{KKsum}, we have included the $\kappa^2$ 
(suppression as a result of gravity coupling), which on summation over the 
high multiplicity of KK modes compensates the suppression in the ADD model.  
In this analysis we have followed the convention of \cite{HLZ}.

Improved theoretical predictions to higher orders in QCD have now been 
performed for cross sections of pair production processes {\em viz.}\ 
di-lepton \cite{dilepton}, di-gauge boson ($\gamma\gamma$ \cite{diph1}, 
$ZZ$ and $W^+ W^-$ \cite{diZZ}), which in extra dimension models
could result from the exchange of a virtual KK mode in addition to the 
usual SM contribution.  The real emission of KK modes lead to large 
missing $E_T$ signals {\em viz.}\ mono-jet \cite{jEt}, mono-photon 
\cite{phEt}, mono-Z boson, and mono-$W^\pm$ boson \cite{ZWEt}.  The 
next to leading order (NLO) QCD corrections in some of the above processes 
are quite substantial and their inclusion in the computation also lead to a
reduction of theoretical uncertainties, making it possible for the experiments 
to put more stringent bounds on the extra dimension model parameters.

The diphoton final state is an important signal for extra dimension 
searches, as the branching ratio of a KK mode decay to diphoton is 
twice than that of a decay to individual charged lepton pair.  
The quantitative impacts of the NLO QCD correction to the diphoton final
state for extra dimension searches have been studied in \cite{diph1},
where various infrared safe observable were studied using phase
space slicing method.  The factorisation scale dependence gets reduced
when ${\cal O} (\alpha_s)$ corrections are included.  Fixed order
calculation truncated to NLO, at best yields results for sufficiently
inclusive observable.  Combining fixed order NLO and parton shower Monte
Carlo (PS) \cite{mc@nlo,powheg}, would extend the coverage of the
kinematical region to consistently include resummation in the 
collinear limit and also make a more exclusive description of the final 
state and get as realistic as possible to the experimental situation.
The flexibility to incorporate hadronisation models and capabilities 
to simulate realistic final state configurations that can undergo 
detector simulations are the main advantage for the experimental
collaborations.

ATLAS \cite{atlas} and CMS \cite{cms} have analysed the diphoton invariant
mass spectrum,  using a constant K-factor for the full range of the invariant
mass distribution to put lower bounds on extra dimension scale to NLO
accuracy.  However, this choice is not sensitive to possible distortions of 
distributions that can arise at NLO.
Our present analysis will
further help to put more stringent bounds on the model
parameters.  In this analysis we have considered various distributions
for the ADD model parameters $d=$ 2 to 6 with appropriate $M_S$ value as
bounded by the experiments \cite{atlas,cms}.  In Table \ref{table_Ms}, the 
$M_S$ values for different extra dimensions $d$ used in this analysis have 
been tabulated.
\begin{table}[tbh]
\begin{center}
 \begin{tabular}{|c|c|c|c|c|c|}
  \hline
$d$ &2 &3 &4 &5 &6 \\ \hline
 $M_S$ (TeV) &3.7 &3.8 &3.2 &2.9 &2.7 \\ \hline
\end{tabular}
\end{center}
\caption{\label{table_Ms}
$M_S$ values used for the various extra dimensions $d$ in our analysis.
}
\end{table}
For relevant observables we consider the fixed order results to NLO accuracy 
and include PS.  Factorisation, renormalisation scale uncertainties and PDF
uncertainties are also estimated in an automated way \cite{scale}.  For 
photon isolation, both smooth cone isolation and the experimental isolation 
criteria are considered.

The rest of the paper is organised as follows.  In section 2, we discuss the 
NLO results to the diphoton final state in the ADD model and the essential 
steps needed to implement the parton showering to NLO accuracy.  
In section 3, we present selected numerical results and estimate the various 
theoretical uncertainties. Finally, in section 4 our conclusions are
presented.

\section{NLO + parton shower} 

Since the KK modes couple universally to the SM particles through the energy
momentum tensor, both the $q \bar q$ and $gg$ channel would contribute to
the diphoton final state at leading order (LO).  In the SM, the $gg$ channel
starts only at next to next to leading order (NNLO) {\it via} the finite box
contribution through quark loop and the large gluon-gluon flux at the LHC 
makes this contribution potentially comparable to the LO results.  In the 
invariant mass region of interest to extra dimension searches, the box diagram 
contribution is not significant enough \cite{diph1}.

All the partonic contributions to NLO in QCD have been calculated for the 
diphoton final state \cite{diph1}, for both large (ADD) \cite{ADD} and 
warped (RS1) \cite{RS1} extra dimension models.  QCD radiative corrections 
through virtual one loop gluon and real emission of gluons to the 
$q ~\bar q \to \gamma ~\gamma $ subprocess, would contribute to both SM and 
extra dimension models.  The $q (\bar q) ~g \to q (\bar q) ~\gamma ~\gamma $ 
begins to contribute for both SM and extra dimension models at NLO.  The LO 
$g ~g \to \gamma ~\gamma$ extra dimension process will also get one loop 
virtual gluon and real gluon emission radiative corrections.  There will 
also be interference between the SM and extra dimension model to give 
contributions up to ${\cal O} (\alpha_s)$.  In this analysis we have not 
included the ${\cal O} (\alpha_s)$ corrections as a result of the interference 
between the SM box diagram contribution and LO extra dimension contribution 
to the $g ~g \to \gamma ~\gamma$ subprocess, as it is quite suppressed in the 
region of interest to extra dimension models.

The $q (\bar q) ~g \to q (\bar q) ~\gamma ~\gamma $ NLO contribution 
has an additional QED collinear singularity when the photon gets 
collinear to the emitting quark and can be absorbed into the 
fragmentation function 
which gives the probability of a parton fragmenting into a photon.
Parton fragmentation functions are additional non perturbative inputs 
which are not very well known.  At the LHC, secondary photons
as a result of hadron decaying into collinear photons and jets faking
as photon are taken care of by photon isolation criteria \cite{atlas,cms} 
which also substantially reduces the fragmentation contribution.    
Since the fragmentation is essentially a collinear effect, the 
fragmentation function can be avoided by the smooth cone isolation 
proposed  by Frixione \cite{frixione}, which ensures that in no 
region of the phase space the soft radiation is eliminated.  The smooth cone 
isolation is able to eliminate the not so well known fragmentation 
contribution and at the same time, ensures infrared safe (IR) observable.  
Centered in the direction of the photon in the pseudo rapidity $(\eta)$
and azimuthal angle $(\phi)$ plane, a cone of radius 
$r=\sqrt{(\eta-\eta_\gamma)^2 + (\phi-\phi_\gamma)^2}$ is defined.
The hadronic activity $H(r)$ is defined as the sum of hadronic transverse 
energy in a circle of radius $r< r_0$ and $E_T^{\gamma}$ is the transverse
energy of the photon.  For all cones with $r \le r_0$ the isolation 
criterion $H(r) < H(r)_{\rm max}$ has to be satisfied, where $H(r)_{\rm max}$
is defined as 
\begin{eqnarray}
H(r)_{\rm max} = \epsilon_\gamma ~E_T^{\gamma} \left (\frac{1-\cos r }{1-\cos r_0} 
\right)^n ~.  
\label{frix_iso}
\end{eqnarray}
Efforts for the experimental implementation of the smooth cone 
isolation is on going.

Automation is an essential ingredient of this work.  We have chosen to work in 
the aMC@NLO framework \cite{scale2}, which automatises the MC@NLO formalism 
\cite{mc@nlo} to match NLO computations with parton showers. In this paper we 
present results matched to HERWIG \cite{herwig}.  For the NLO computation, 
isolation of IR poles and phase space integration are carried out by MadFKS
\cite{madFKS}, which automatises the FKS subtraction method \cite{FKS} using 
the MadGraph \cite{madgraph} matrix-element generator, whereas for one-loop amplitudes 
the results of Ref.\ \cite{diph1} are used.  The automation within the MadGraph 
framework requires a new HELAS \cite{helas} subroutine to 
calculate helicity amplitudes with massive 
spin-2 particles \cite{hagi1,hagi2}.  In addition, for our present analysis we
have implemented the sum over the KK modes to take care of the virtual KK mode 
sum (Eq.\ \ref{KKsum}) that contributes to process in the ADD model \cite{hagi2}.  
We use this framework to generate the events for 8 TeV run at the LHC.  For the 
invariant mass 
distributions we have reproduced the results of \cite{diph1} using the fixed order 
results from MadFKS.  Also numerical cancellation of the singularities from the real 
and virtual terms have been explicitly checked.  

\begin{figure}
\centerline{ 
\includegraphics[width=13cm]{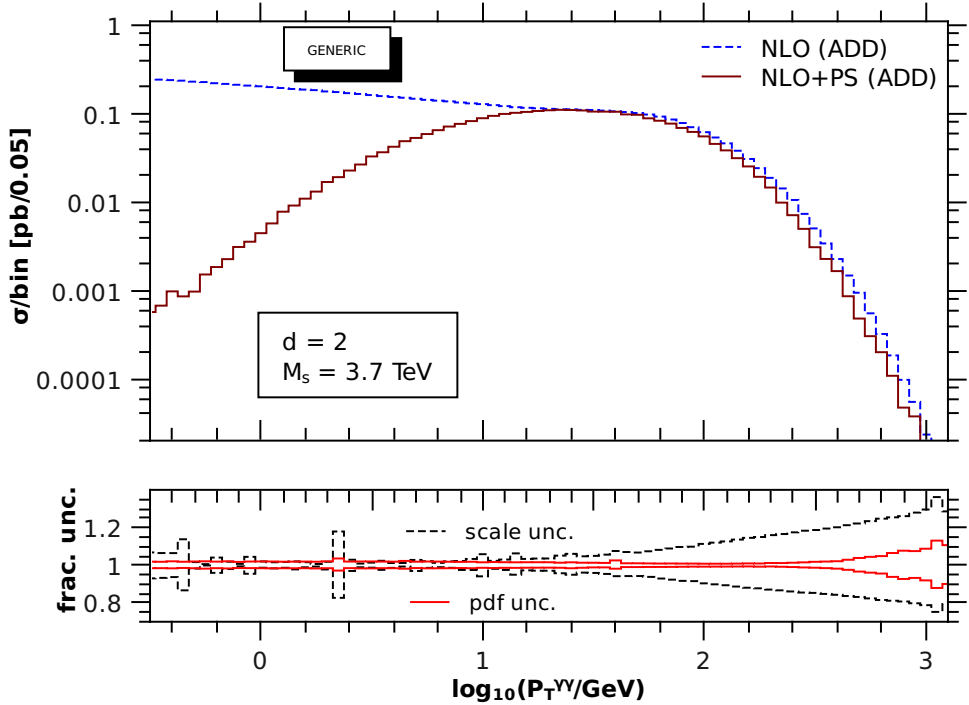}
}
\caption{\label{fig_pt} Transverse momentum distribution $p_T^{\gamma \gamma}$ 
of the diphoton for the fixed order NLO and NLO+PS.  The ADD model parameters
used are $d=2$ and $M_S=3.7$ TeV.  The lower inset displays the scale and PDF, 
fractional uncertainties for the NLO+PS results.
}
\end{figure}

\begin{figure}
\centerline{
\includegraphics[width=8cm]{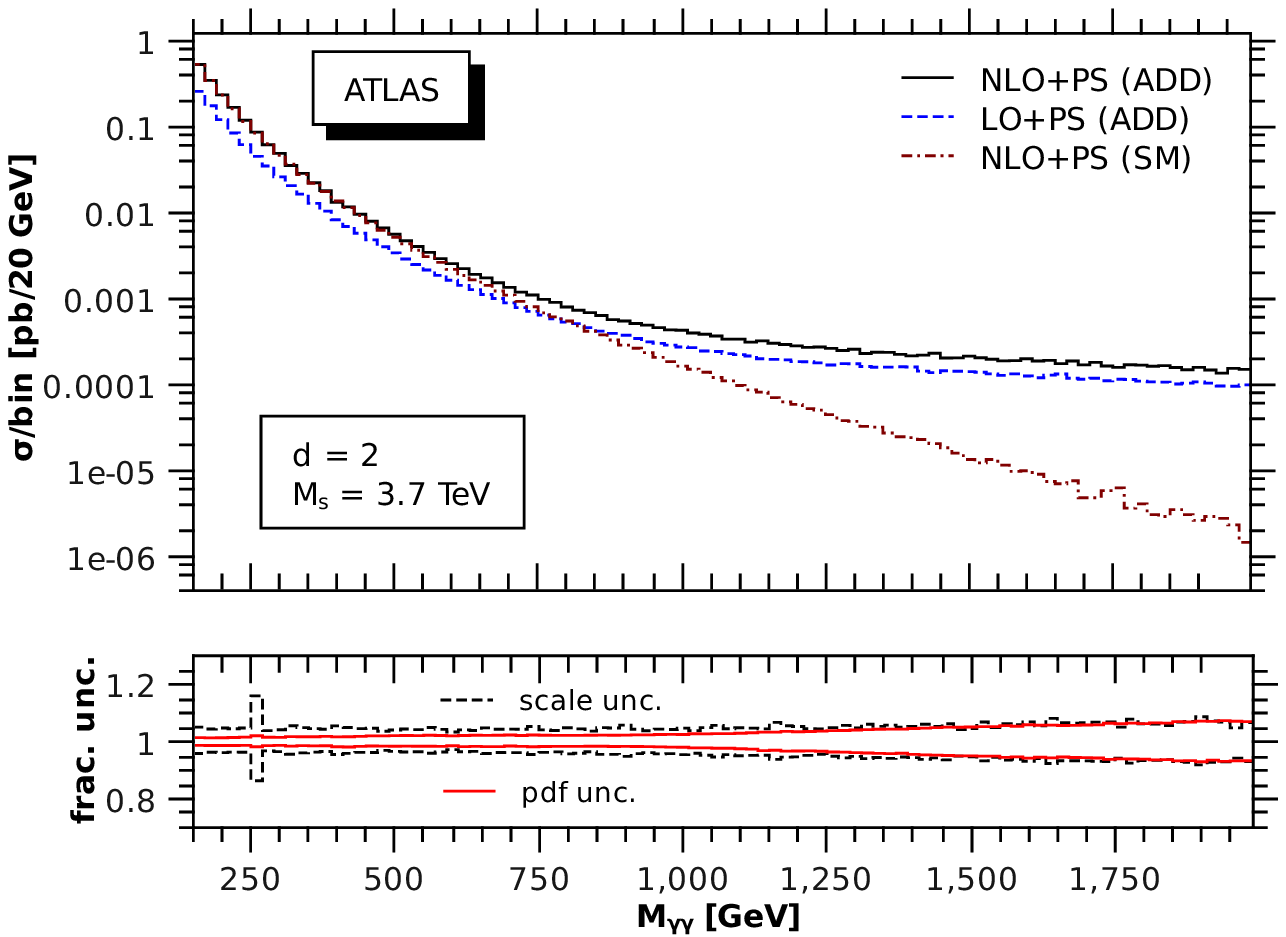}
\includegraphics[width=8cm]{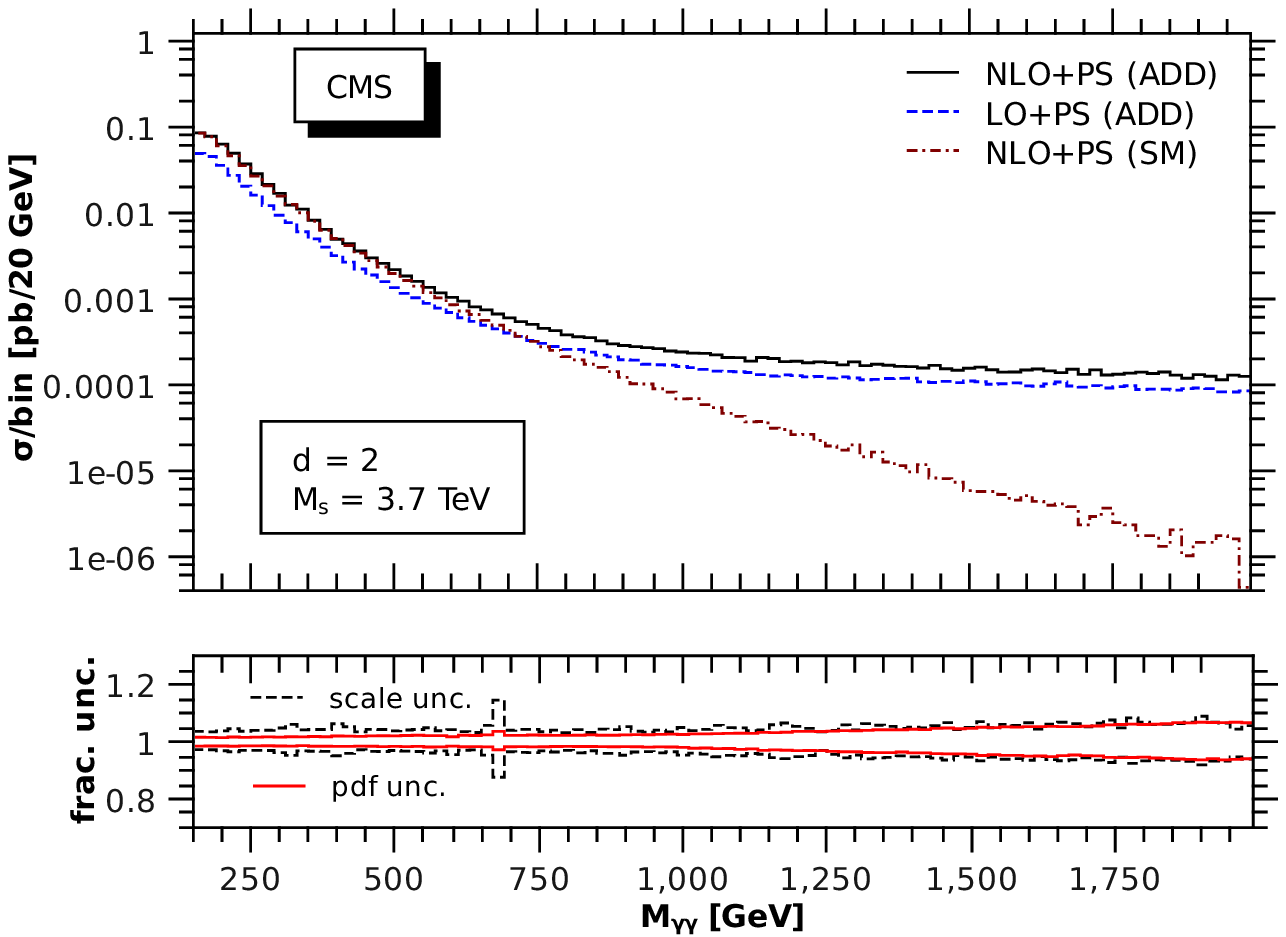}}
\caption{\label{fig_Q} 
Invariant mass distribution $M_{\gamma \gamma}$ for ATLAS (left panel) and
CMS (right panel) for $d=2$ extra dimensions and $M_S = 3.7$ TeV.  The SM
contribution to NLO+PS and ADD to LO+PS and NLO+PS have been plotted.  For
the NLO+PS (ADD) results, the lower inset displays the scale and PDF fractional 
uncertainties.
}
\end{figure}

\section{Numerical results}

In this section, we present the results for various kinematic distributions 
of photon pair in SM and ADD model.  We have included all the subprocess 
contributions to NLO.  The following input parameters are used: 
$\alpha_{em}^{-1}=132.507$, $G_F=1.16639 \times 10^{-5}$ GeV$^{-2}$, $m_Z=91.188$ 
GeV and 
MSTW2008(n)lo68cl for (N)LO parton distribution functions (PDF)  \cite{mstw}.
The MSTW PDF also sets the value of the strong coupling $\alpha_s (m_Z)$ at 
LO and NLO in QCD.
The renormalisation and factorisation scales are chosen as
$\mu_F = \mu_R = M_{\gamma \gamma}$, the invariant mass of the photon pair.  
The events that have to be showered are generated using
the following generation cuts:
$|\eta_{\gamma_{1,2}} | < 2.6$, $p_T^{\gamma_{1,2}} >20$ GeV, diphoton 
invariant mass $100 ~{\rm GeV} < M_{\gamma \gamma} < M_S$ and the photon 
isolation is done using the Frixione isolation with 
$r_0 = 0.38$, $\epsilon_\gamma =1$ and $n=2$ (see Eq.\ (\ref{frix_iso})).
More specific analysis cuts can be applied subsequently to 
generate the events.

The dependence of the prediction of an observable on the factorisation and 
renormalisation scales, is a result of the uncalculated higher order contributions,
which can be estimated by varying $\mu_F$ and $\mu_R$ independently around the
central value $\mu_F = \mu_R = M_{\gamma\gamma}$.  The variation is done by the 
following assignment
$\mu_F = \xi_F ~ M_{\gamma\gamma}$
and
$\mu_R = \xi_R ~ M_{\gamma\gamma}$,
where the values for $(\xi_F,\xi_R)$ used are (1,1), (1/2,1/2), (1/2,1), (1,1/2),
(1,2), (2,1), (2,2).  The various ratios of $\mu_F$, $\mu_R$ and $M_{\gamma\gamma}$
that appear as arguments of logarithms in the perturbative expansion to NLO are
within the range [1/2,2].  The variation of both $\mu_F$ and $\mu_R$ are taken as the
envelope of the above individual variations.  Variation of only $\mu_F$
would involve the choice $\xi_R=1$ and varying $\xi_F$ and vice-versa for variation 
of only $\mu_R$.  The PDF uncertainties are estimated using the prescription given 
by MSTW \cite{mstw}.  Fractional uncertainty defined as the ratio of the variation 
about the central value divided by the central value, is a good indicator of the 
scale and PDF uncertainties and is plotted in the lower insets to the various 
figures. As described in Ref.~\cite{scale}, the generation of these uncertainty bands
can be done at virtually no extra CPU cost within the aMC@NLO framework.

To begin with, we compare the fixed order NLO result with NLO+PS for the 
transverse momentum of the diphoton $\log_{10} p_T^{\gamma\gamma}$ using
generic cuts: 
$M_{\gamma\gamma} >140$ GeV, $|\eta_\gamma | < 2.5$, $p_T^{\gamma_1} >40$ GeV, 
$p_T^{\gamma_2} >25$ GeV and $r_0=0.4$.
In Fig.\ \ref{fig_pt}, $\log_{10} p_T^{\gamma\gamma}$ distribution is plotted 
for $d=2$ with appropriate $M_S$ value.  It is clear that at low 
$p_T^{\gamma \gamma}$ 
values, NLO+PS correctly resums the Sudakov logarithms, leading to a suppression
of the cross section, while the fixed order NLO results diverges for
$p_T^{\gamma \gamma} \to 0$.  At high $p_T^{\gamma \gamma}$, the NLO 
fixed order and NLO+PS results are in agreement. In the lower inset of the 
Fig.\ \ref{fig_pt}, we have the scale and PDF variation of the NLO+PS, which   
increase with $p_T^{\gamma\gamma}$ as observed in \cite{scale1}.

\begin{figure}
\centerline{ 
\includegraphics[width=8cm]{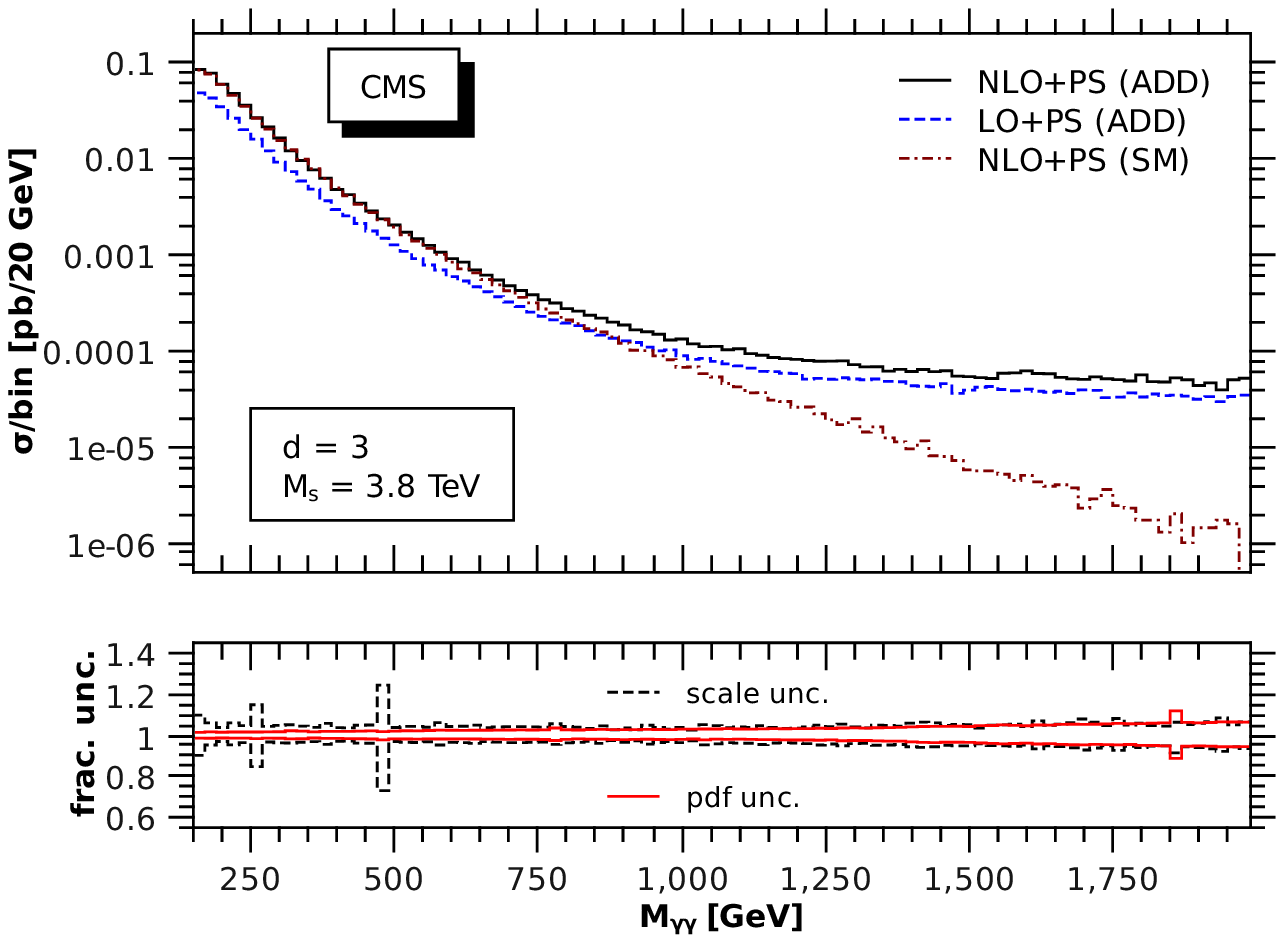}
\includegraphics[width=8cm]{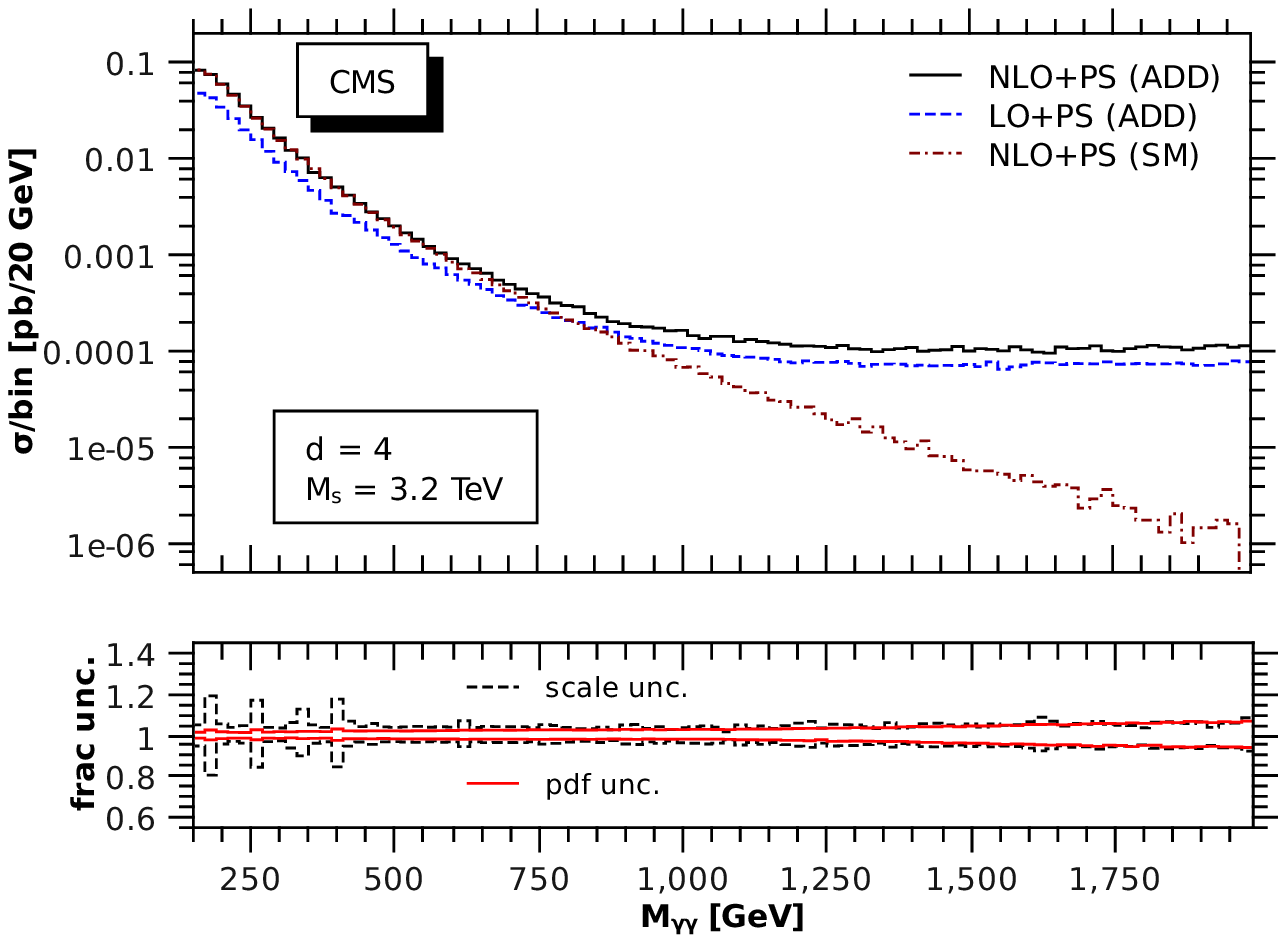}
}
\caption{\label{fig_Q34}
Invariant mass distribution for $d=3$ (left panel) and $d=4$ (right panel) 
is plotted for ADD and SM contributions to NLO+PS accuracy.  The lower 
insets gives the corresponding scale and PDF, fractional uncertainties for 
NLO+PS (ADD).  
}
\end{figure}

We now present the results for the various kinematical distributions to NLO 
accuracy with PS (labelled as NLO+PS), for analysis specific cuts.  Both the 
experiments ATLAS and CMS have looked for diphoton invariant mass in the 
region 140 GeV $< M_{\gamma \gamma} < M_S$.  
ATLAS cuts \cite{atlas}:
the rapidity of the individual photons are in the region $|\eta_\gamma| < 2.37$, 
with an exclusion region $1.37 <|\eta_\gamma| < 1.52$, the transverse momentum 
of the individual photons $p_T^\gamma >25$ GeV and for photon isolation: 
sum of transverse energy of hadrons $\sum E_T (H) <5$ GeV with $\Delta r < 0.4$.  
$\Delta r=\sqrt{\Delta \phi^2 +\Delta \eta^2}$ is
a cone in the azimuthal angle, rapidity plane.
For CMS the corresponding cuts are \cite{cms}:
$|\eta_\gamma| < 1.44$, 
$p_T^\gamma >70$ GeV, 
photon isolation: 
(i) sum of the energy of hadrons $\sum E (H) < 0.05 E^\gamma$ with $\Delta r < 0.15$, 
(ii) sum of transverse energy of hadrons $\sum E_T (H) <2.2 ~{\rm GeV} +0.0025 ~E_T^\gamma$ 
    with $0.15 <\Delta r < 0.4$.  
In addition to the ATLAS and CMS photon isolation, if we also include the
Frixione isolation criteria, there is no appreciable change in the results.

\begin{figure}
\centerline{ 
\includegraphics[width=5.2cm]{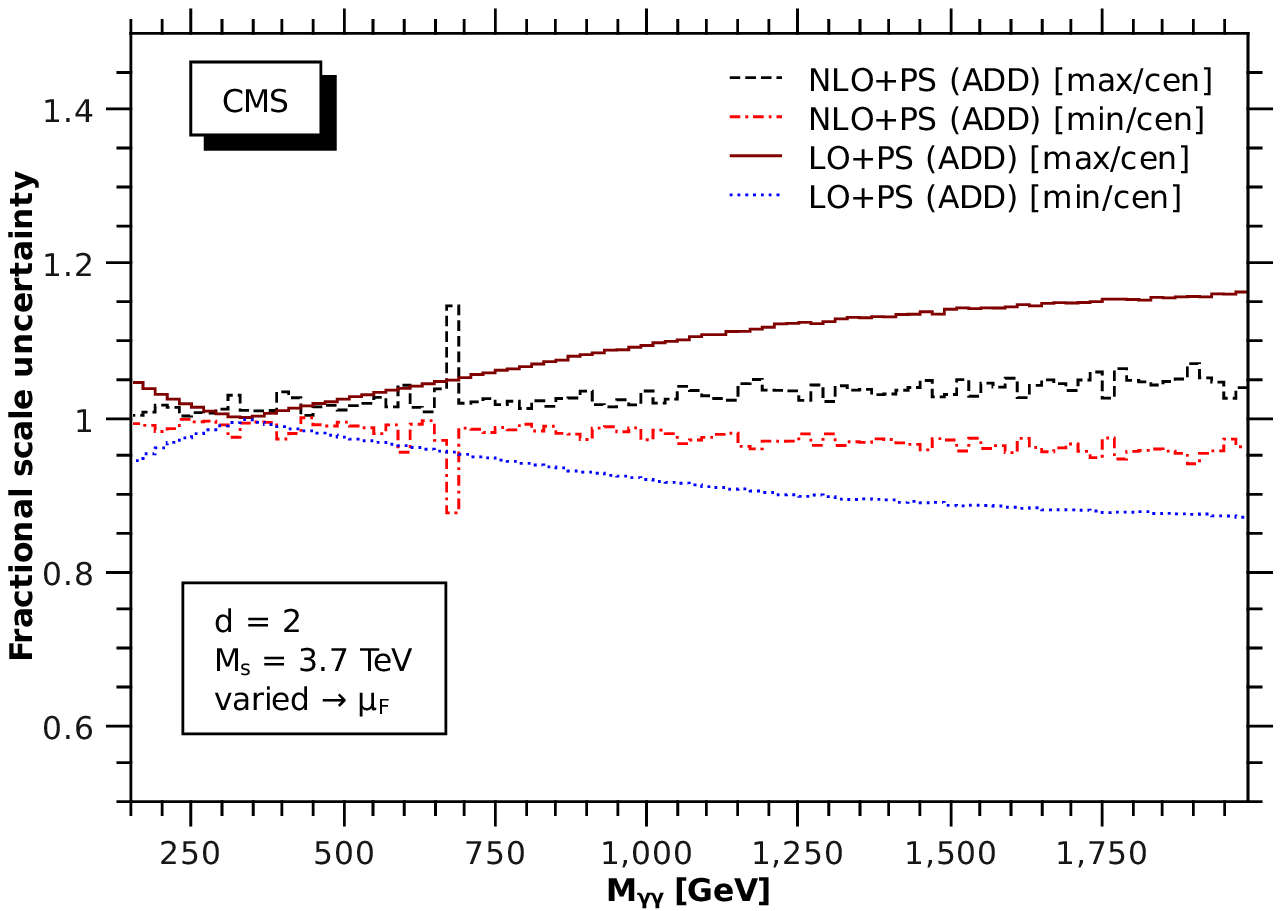}
\includegraphics[width=5.2cm]{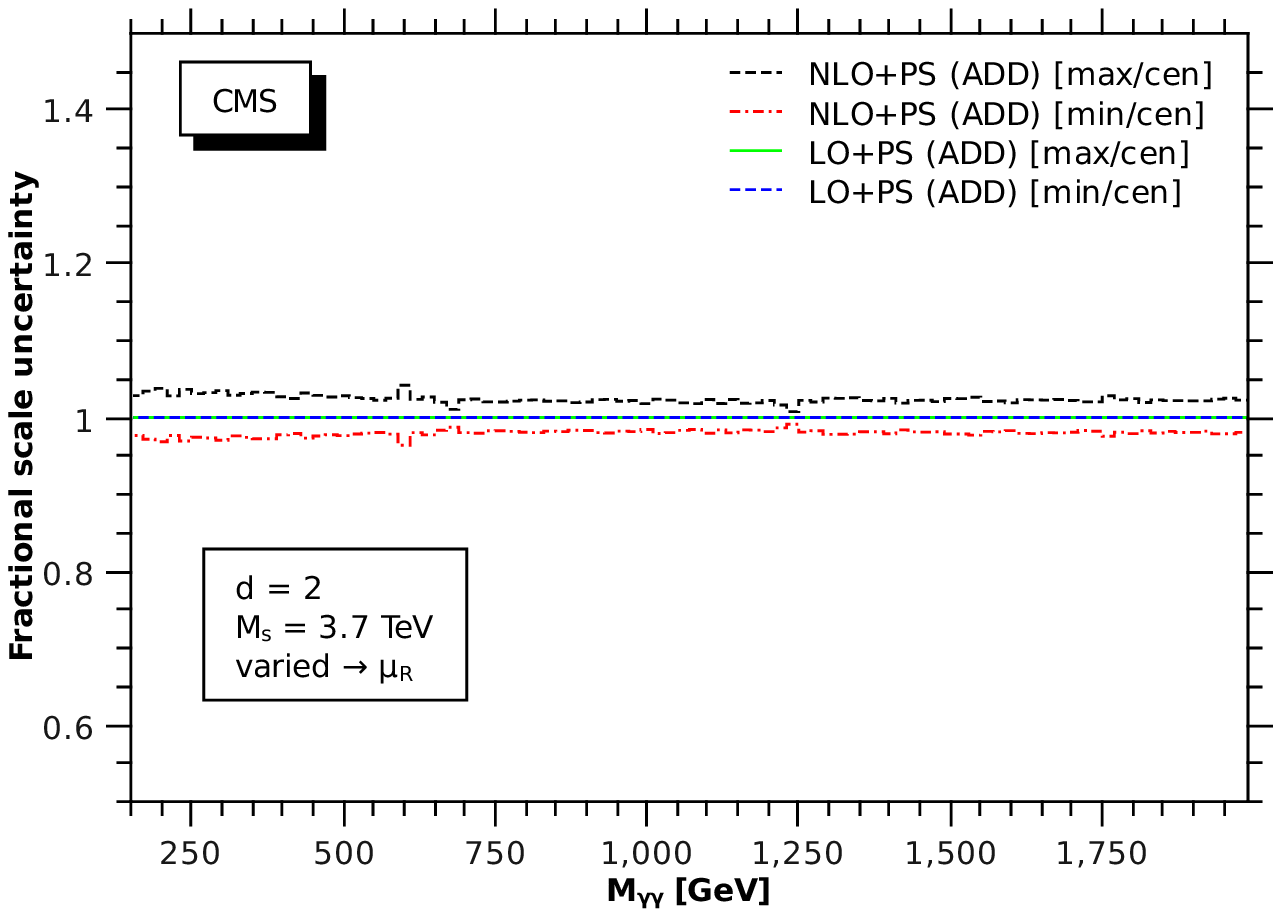}
\includegraphics[width=5.2cm]{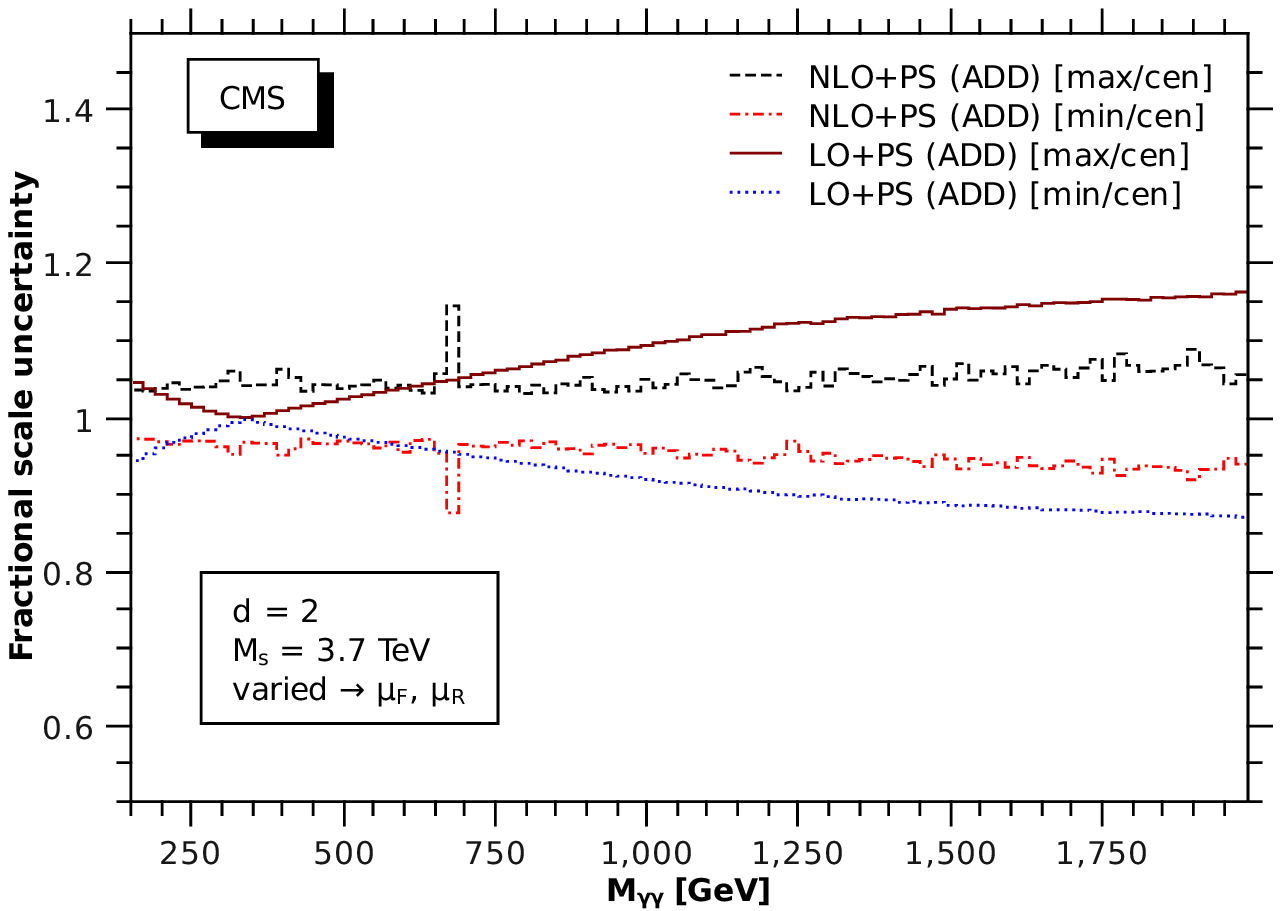}}
\caption{\label{fig_scale}
For the invariant mass distribution, with $d=2$ and $M_S=3.7$ TeV,
the fractional uncertainties as a result of $\mu_F$ variation (left panel),
$\mu_R$ variation (central panel) and $\mu_F, ~\mu_R$ variation (right panel).
}
\end{figure}

In Fig.\ \ref{fig_Q}, we have plotted invariant mass distributions $d\sigma/
d M_{\gamma\gamma}$ of photon pair in the SM as well as in the ADD model for ATLAS 
(left panel) and CMS (right panel).  For ADD model we have obtained the 
distributions for $M_S=3.7$ TeV and $d=2$.  The central value curves correspond
to the choice $\mu_F=\mu_R = M_{\gamma\gamma}$, have been plotted for the ADD 
(NLO+PS) and purely SM (NLO+PS) contribution.  The label ADD refers to the total
contribution coming from SM, ADD and the interference between them.  The corresponding ADD 
(LO+PS) contribution gives an indication of the quantitative impact of the NLO 
QCD correction.  At larger invariant mass of the photon pair the ADD
effect is dominant.  To demonstrate the sensitivity of our predictions to 
the choice of scale and PDF uncertainties, in the lower inset fractional
uncertainty by varying (a) both $\mu_F$ and $\mu_R$ and (b) PDF error sets, are 
plotted.
The difference in the distribution in Fig.\ \ref{fig_Q}
for ATLAS and CMS can be attributed to the very different cuts used for their
analysis.
In Fig.\ \ref{fig_Q34}, the corresponding plots for 
$d=3,4$ are plotted for the CMS cuts.  The choice of $M_S$ used for the 
plots corresponds to the lower bounds obtained by \cite{atlas,cms} using
the diphoton process.
By including higher order corrections, the scale dependence goes down from about 25\% at 
LO, to about 10\% at NLO, as can be seen from the ratio plots.  The PDF uncertainty does 
not change significantly and remains about 8\%.

\begin{figure}
\centerline{
\includegraphics[width=8cm]{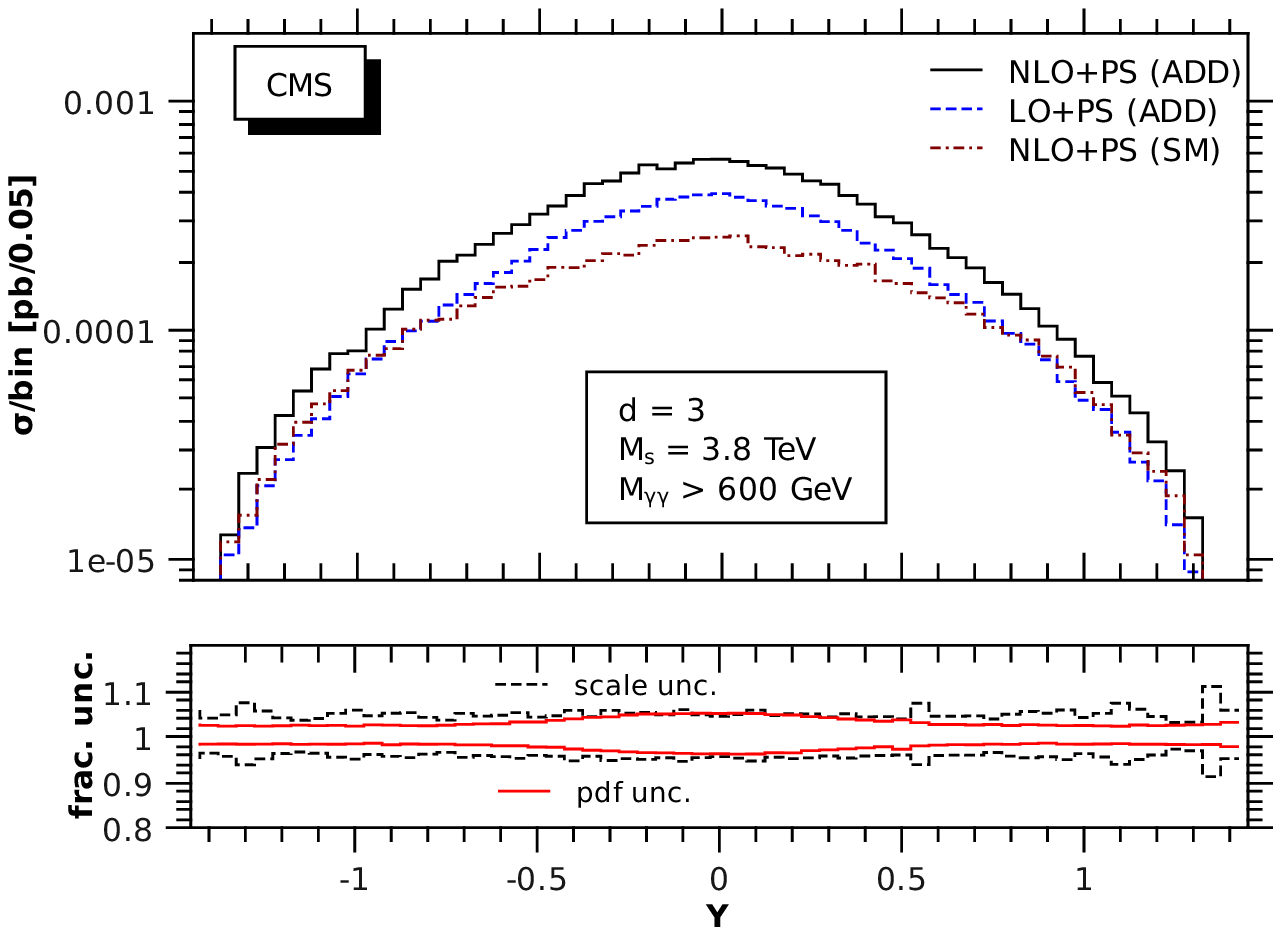}
\includegraphics[width=8cm]{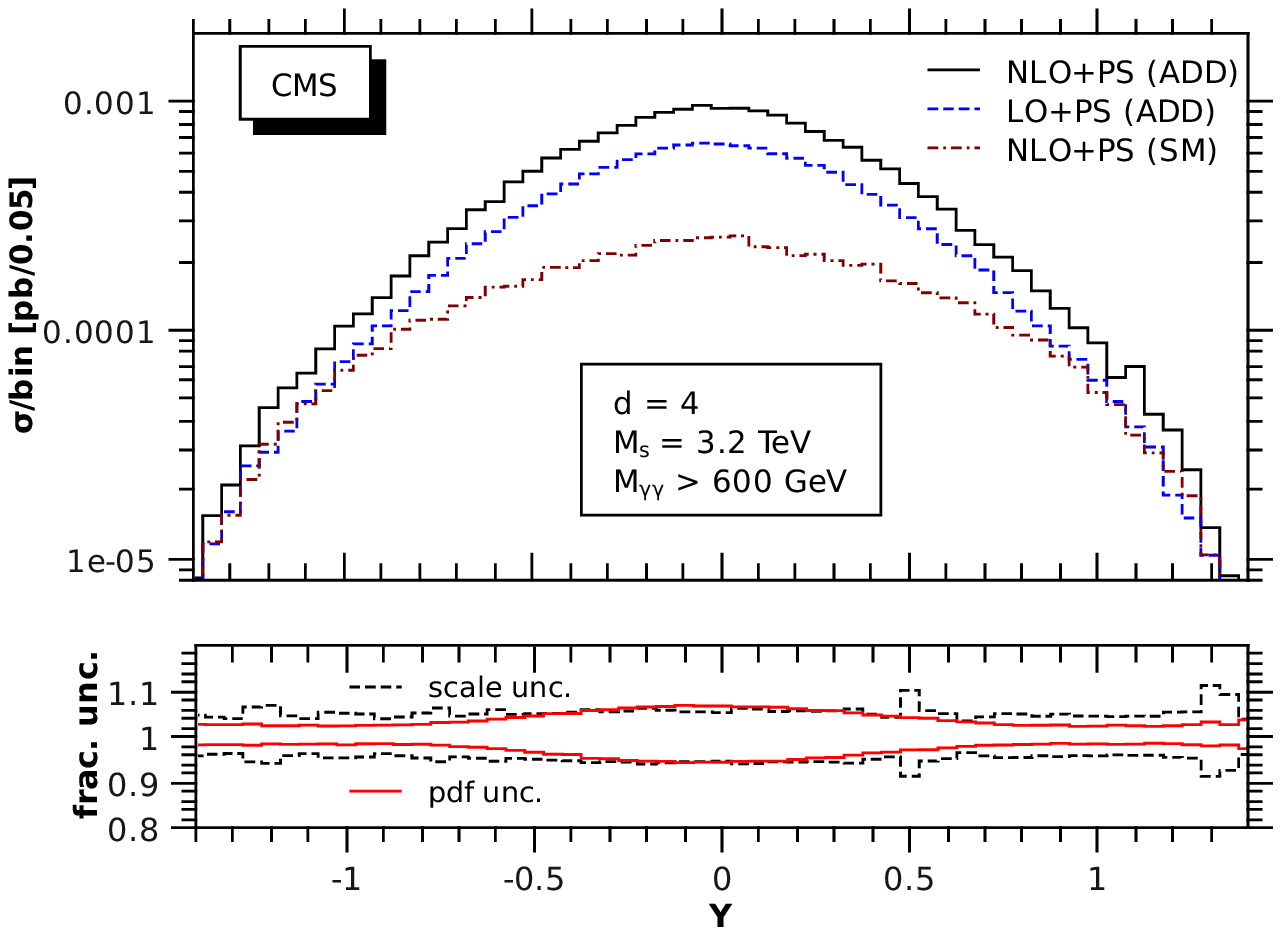}
}
\caption{\label{fig_Y}
The rapidity $(Y)$ distribution of the diphoton pair for $d=3$ (left panel) and $d=4$ 
(right panel) for SM (NLO+PS) and ADD (LO+PS and NLO+PS).  The lower inset displays 
the corresponding fractional scale and PDF uncertainties of the NLO+PS (ADD) results.
}
\end{figure}

We now consider the fractional scale uncertainties on the invariant mass 
distribution as a result of the variation of the scales $\mu_F$ and $\mu_R$ 
(both independently and simultaneously) in going from LO+PS to NLO+PS.  Note 
that the LO cross sections depend only on $\mu_F$ through the PDF sets, but 
at NLO level the scale $\mu_R$ enters through $\alpha_s(\mu_R)$ and 
$\log(\mu_F/\mu_R)$ 
coming from the partonic cross sections after mass factorisation.  As expected
the inclusion of NLO QCD correction reduces the factorisation scale dependence
resulting from the LO observable which is clear from Fig.\ \ref{fig_scale} 
(left panel).  In the high $M_{\gamma\gamma}$ region, the uncertainty of about
25\% at LO+PS gets reduced to 5\% when NLO+PS corrections are included.
On the other hand, the $\mu_R$ dependence enters only at NLO 
level (see middle panel of Fig.\ \ref{fig_scale}) which will get reduced only 
if NNLO corrections are included.  Hence, we see our NLO corrections are 
sensitive to the choice of $\mu_R$ but the variation is only 5\% and
is fairly constant for the range of invariant mass considered.
If we vary both $\mu_F$ and $\mu_R$ simultaneously as shown in Fig.\ 
\ref{fig_scale} (right panel), we find that the reduction in the $\mu_F$ scale 
dependence at NLO level is mildly affected by the $\mu_R$ variation in the large 
invariant mass region.  In the small invariant mass region, the LO and NLO results
exhibit smaller $\mu_F$ dependence compared to the large invariant mass region.
But $\mu_R$ dependence coming from the NLO results does not change much with 
the invariant mass $M_{\gamma\gamma}$.  Hence variation due to $\mu_R$ at small 
$M_{\gamma\gamma}$ is larger compared to that resulting from $\mu_F$.  This
explains the behavior at small invariant mass regions where the NLO+PS variation 
is in excess of the LO+PS (see right panel of Fig.\ \ref{fig_scale}).  

\begin{figure}
\centerline{ 
\includegraphics[width=8cm]{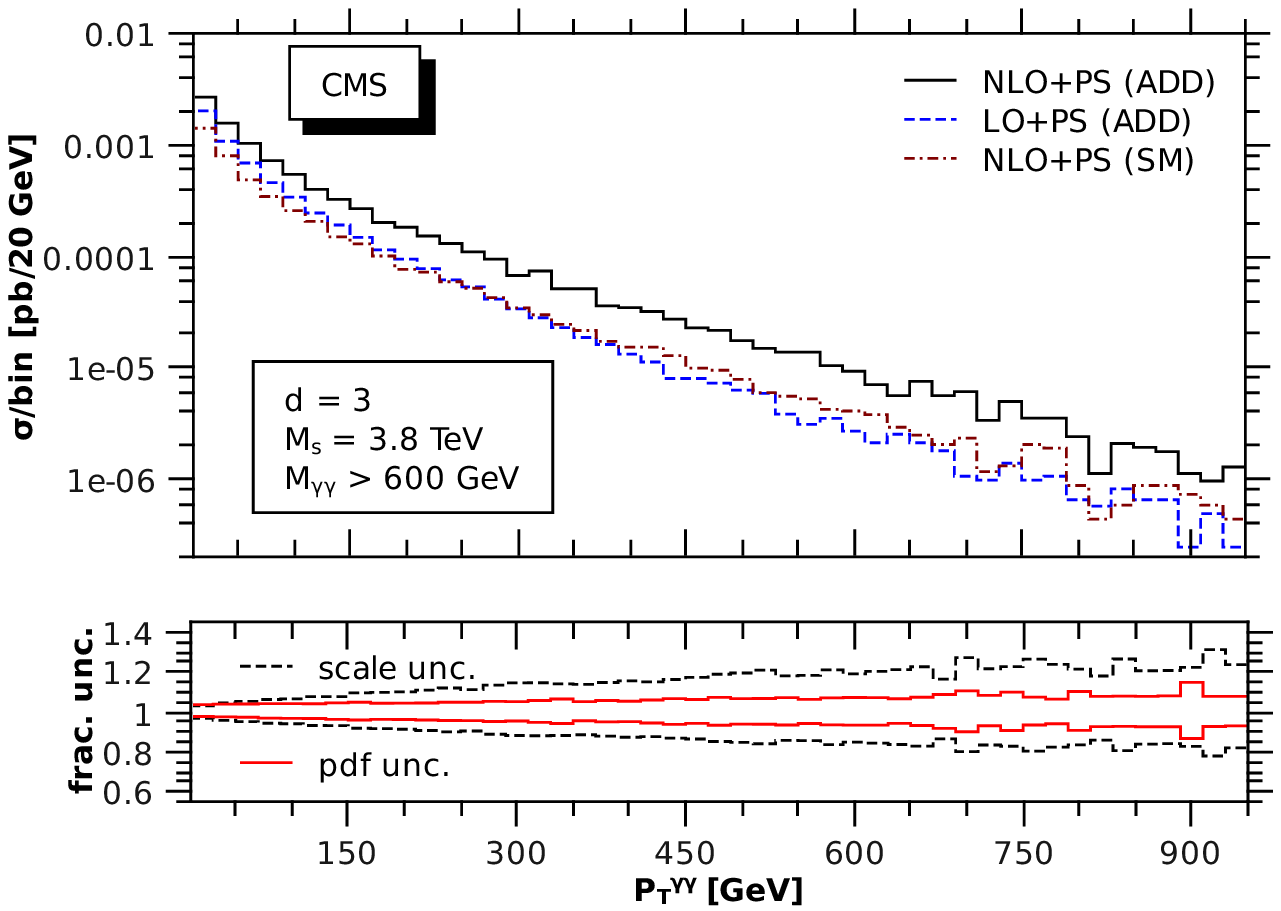}
\includegraphics[width=8cm]{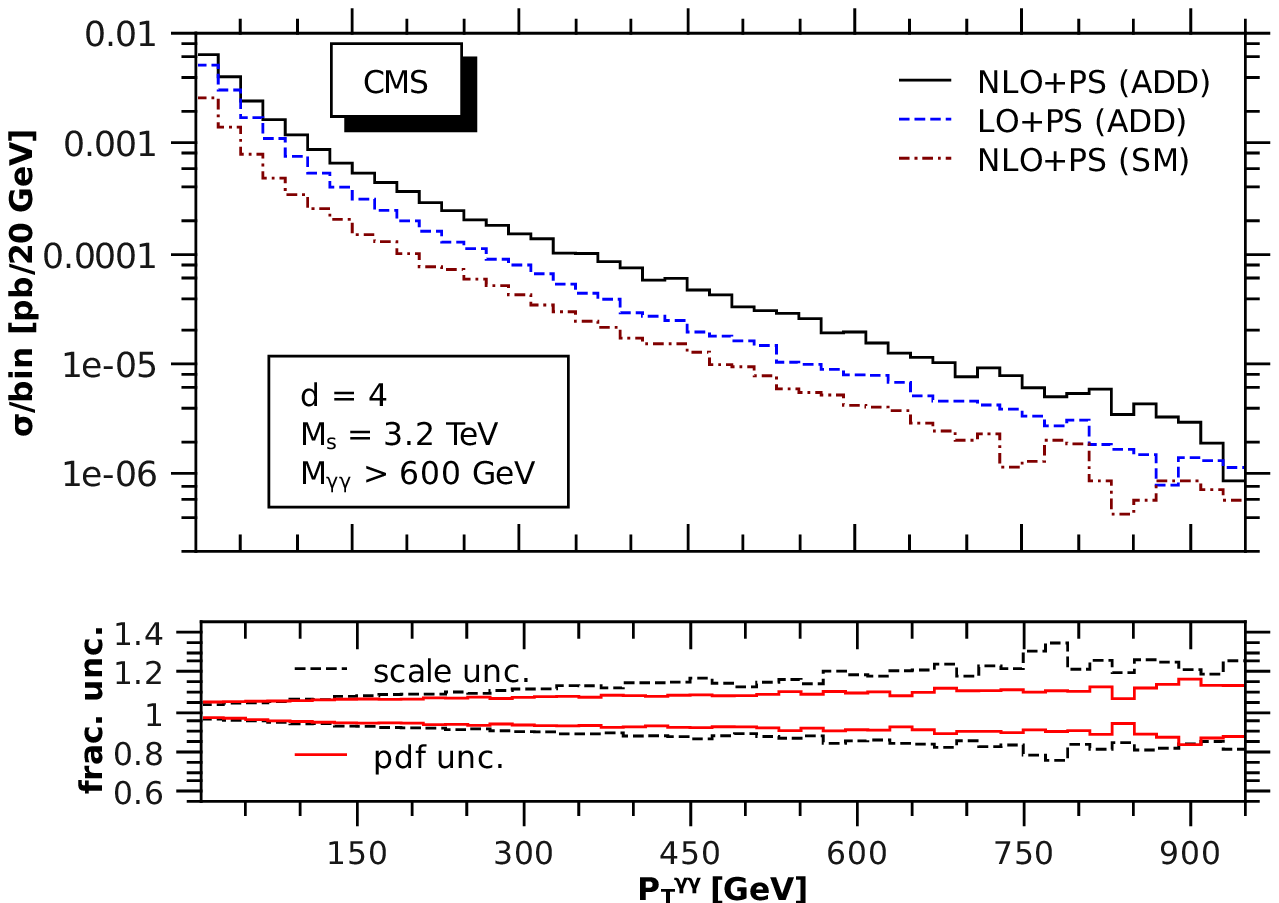}
}
\caption{\label{fig_pt1}
The transverse momentum distribution $p_T^{\gamma\gamma}$ of the diphoton for $d=3$ 
(left panel) and $d=4$ (right panel).
}
\end{figure}

The rapidity distribution of the diphoton pair is plotted in Fig.\ \ref{fig_Y} 
for $d=3$ (left panel) and $d=4$ (right panel).  For this analysis we have 
chosen $M_{\gamma\gamma} >600$ GeV, the region where the effects of ADD model 
begins to dominate over the SM diphoton signal at NLO (see Fig.\ \ref{fig_Q34}).
The scale and PDF uncertainties to NLO are displayed as insets at the bottom of 
each figure.  The scale uncertainties are usually larger than the PDF 
uncertainties in the rapidity distribution except for the central rapidity
region where they are comparable.  For $d=3$ the scale uncertainties are about 
20\% around the central rapidity region, which come down to about 10\% when 
NLO+PS corrections are included.  The PDF uncertainties for LO+PS and NLO+PS 
are comparable.  

Finally, we plot the transverse momentum distribution in Fig.\ \ref{fig_pt1} 
for $d=3$ (left panel) and $d=4$ (right panel), for the SM and ADD model
to NLO+PS accuracy, with $M_{\gamma\gamma}>600$ GeV.  
The ADD results are also plotted for LO+PS.
The scale and PDF uncertainties are displayed as insets at 
the bottom of the plots for NLO+PS (ADD).  

\section{Conclusion}

In this analysis, we have presented the diphoton final state in the large 
extra dimension model to NLO in QCD and matching to parton shower is
implemented using the aMC@NLO framework.  All the subprocesses that contribute
to the diphoton final state from both the SM and ADD model are considered
to NLO in QCD.  This is the first time MC@NLO formalism has
been used for a processes in the ADD model and we hope it would significantly 
help extra dimension searches at the LHC to constrain the ADD model
parameters.  Using a set of generic cuts we first demonstrated the importance 
of NLO+PS over the fixed order NLO computations, by considering the 
$p_T^{\gamma\gamma}$ distribution.  We have presented our results for 
various observables {\em viz.}, invariant mass, rapidity and transverse 
momentum of the diphoton, both for the ATLAS and CMS detector specific 
cuts to NLO+PS accuracy.  It is important to note that there is substantial
enhancement of the various distributions due to the inclusion of NLO
corrections and both the theoretical and PDF uncertainties have been
estimated.  There is a significant decrease in theoretical uncertainties 
from over 20\% at LO to about 10\% when NLO corrections are included. 
The results are presented for different 
number of extra spatial dimensions $d=2-6$ and the respective values of 
fundamental scale $M_S$ that have been experimentally bounded.
The event files for $d=2-6$ are available on the website http://amcatnlo.cern.ch
and we are working on making the code that was used to  generate
these events publicly available.

\section*{Acknowledgements}
PM and VR would like to acknowledge M.\ C.\ Kumar and Anurag Tripathi for previous 
collaboration on diphoton parton level results.  
SS would like to acknowledge useful communication with Priscila de Aquino and 
Fabio Maltoni.  SS would like to thank UGC, New Delhi, for financial 
support.  
MZ would like to thank Fabio Maltoni for reading the manuscript and for 
many inspiring discussions.  The work of MZ is supported by the 
4.4517.08 IISN-FNRS convention.

\end{document}